\documentclass[11pt,a4paper]{article}
\usepackage{jheppub}
\pdfoutput=1

\usepackage{epsfig}
\usepackage{latexsym}
\usepackage{booktabs}
\usepackage{color}
\usepackage[dvipsnames]{xcolor}
\usepackage{footnote}
\usepackage{hyperref}
\usepackage{simplewick}
\usepackage[caption=false]{subfig}
\usepackage{bm}

\newcommand\york{Department of Physics and Astronomy, York University, Toronto, Ontario, M3J 1P3, Canada}

\newcommand\bern{Albert Einstein Center, Universit\"at Bern,  CH-3012 Bern,  Switzerland}
\newcommand\cern{Theory Department, CERN, CH-1211 Geneva, Switzerland}
\newcommand\nycu{Institute of Physics, National Yang Ming Chiao Tung University, 30010 Hsinchu, Taiwan}

\newcommand\kmyork{Department of Mathematics and Statistics, York University, Toronto, Ontario M3J 1P3, Canada}

\newcommand\adelaide{CSSM, University of Adelaide, Adelaide, SA, 5005, Australia}

\makeatletter
\gdef\@fpheader{}
\makeatother

\begin{document}

\title{Diquark properties from full QCD lattice simulations}

\dedicated{\vspace{5ex} Dedicated to the memory of Artan Bori\c{c}i.}

\author{Anthony~Francis$^{a,b,c}$,\; Philippe~de~Forcrand$^b$,\; Randy~Lewis$^d$ \;and Kim~Maltman$^{e,f}$}

\affiliation{$^a$\bern,\\ $^b$\cern,\\ $^c$\nycu,\\ $^d$\york,\\ $^e$\kmyork,\\ $^f$\adelaide }
\date{\today}

\preprint{CERN-TH-2021-093}

\abstract{
We study diquarks on the lattice in the background of a 
static quark, in a gauge-invariant formalism with quark masses 
down to almost physical $m_\pi$. We determine mass differences 
between diquark channels as well as 
diquark-quark mass differences. The lightest 
and next-to-lightest diquarks have
``good'' scalar, $\bar{3}_F$, $\bar{3}_c$, $J^P=0^+$,
and ``bad'' axial vector, $6_F$, $\bar{3}_c$, $J^P=1^+$, 
quantum numbers, and a bad-good mass difference for $ud$ flavors, $198(4)~\rm{MeV}$,
in excellent agreement with phenomenological 
determinations. Quark-quark attraction is 
found only in the ``good'' diquark channel. 
We extract a corresponding diquark 
size of $\sim 0.6~\rm{fm}$ and perform a 
first exploration of the ``good'' diquark shape,
which is shown to be spherical. Our results provide quantitative support for modeling the 
low-lying baryon spectrum using good light 
diquark effective degrees of freedom.
}

\maketitle

\noindent

\section{Introduction }
The
renewed popularity ({\it e.g.} 
\cite{Jaffe:2004ph,Ali:2017jda,Barabanov:2020jvn})
of the old diquark idea\cite{GellMann:1964nj,Ida:1966ev,Lichtenberg:1982jp} makes a detailed {\it ab initio} lattice
study useful and timely.

Diquarks, like quarks, carry an open 
color index, and hence do not exist as
asymptotic states. Since
contracting a diquark with a quark produces a baryon operator, 
a diquark with an anti-diquark a tetraquark operator, {\it etc.},
effective diquark degrees of freedom may be useful building blocks
for phenomenological descriptions of hadronic states. 
Such a description might prove successful if
diquarks are compact objects with fewer degrees 
of freedom than the quark pairs they represent. The phenomenological 
success of diquark models \cite{Cahill:1987qr,Maris:2002yu,Santopinto:2004hw,Barabanov:2020jvn} supports this possibility, 
provided one assumes the diquarks have ``good'' ($\bar{3}_F$, 
$\bar{3}_c$, $J^P=0^+$) flavor, color and Dirac quantum 
numbers. This assumption is natural 
since both
one-gluon-exchange~\cite{DeRujula:1975qlm,DeGrand:1975cf} and instanton interactions~\cite{tHooft:1976snw,Shuryak:1981ff,Schafer:1996wv} are 
attractive in this channel. The present work aims to investigate this picture quantitatively 
by studying diquark masses,
sizes and spatial correlations using first-principle 
lattice QCD simulations.

Since diquarks are colored, and not gauge-invariant, neither are their properties.
One way to deal with this issue is to
work in a fixed gauge,
typically Landau 
gauge or a variant thereof, see, {\it e.g.}, 
the lattice studies of 
Refs.~\cite{Hess:1998sd,Bi:2015ifa,Babich:2007ah}. 
The drawback is that the resulting
diquark properties
depend on the gauge choice. This problem is
well known for the size determination \cite{Teo:1992zu,Negele:2000uk,Alexandrou:2002nn}, 
though diquark masses, and even 
mass differences, are also affected
since these are extracted from the temporal decay 
rates of appropriate correlators, which will 
change in a gauge
non-local in time 
like Landau gauge. Alternately, one can 
introduce a static color source which, together
with the diquark, forms a color singlet baryon,
whose mass is gauge-invariant. Since the mass 
of such a static-light-light baryon 
diverges in the continuum limit, the quantities 
of interest are mass {\em differences} between
various diquark channels. The diquark 
size can also be obtained in a gauge-invariant 
way, from the spatial decay rate of the quark 
density-density correlator at fixed 
time \cite{Orginos:2005vr,Alexandrou:2005zn,Alexandrou:2006cq,Green:2010vc,Green:2012zza,Fukuda:2017mmh}.

We adopt the second, gauge-invariant approach of \cite{Alexandrou:2005zn,Alexandrou:2006cq,Fukuda:2017mmh}. 
Measurements are taken 
on dynamical $n_f=2+1$, $32^3\times 64$, 
clover-improved Wilson fermion
gauge configurations with lattice spacing 
$a \approx 0.090~\rm{fm}$ 
generated by 
PACS-CS~\cite{Aoki:2008sm,Namekawa:2013vu} and publicly available 
from the JLDG repository~\cite{JLDG}. 
Five ensembles, with pion masses 
$m_\pi =164, 299, 415, 575, 707$ MeV, are 
considered, allowing us to study
the dynamical light-quark mass dependence 
of diquark properties and perform a 
short, controlled extrapolation 
to physical $m_\pi$. We re-use the 
(gauge-fixed, wall-source) quark
propagators from \cite{Francis:2016hui,Francis:2018jyb}.
To connect with previous quenched studies, we also 
employ a new $32^3 \times 64$, $a \approx 0.092~\rm{fm}$,
quenched ensemble with 
valence pion mass $m_\pi=909~\rm{MeV}$.
Static quark propagators are computed using the 
method of 
\cite{DellaMorte:2005nwx,Donnellan:2010mx} with 
HYP1 smearing. See Appendix~\ref{app:latt} for further details.

\section{Diquark spectroscopy }
\label{sec:diqspec}
We first quantify the expected 
reduction in the ``good'' diquark mass by
studying the static-light-light baryon 
spectrum. With $Q$ the static quark, 
$c, C$ denoting charge 
conjugation, and light quarks in a 
$D_\Gamma =q^c C \Gamma q$
diquark configuration, where
$\Gamma$ acts in Dirac space,
we measure the baryon correlators
\begin{align}
    C_{\Gamma}(t)=\sum_{\vec x} \Big\langle [D_\Gamma Q](\vec x,t)~[D_\Gamma Q]^\dagger(\vec 0,0)  \Big\rangle \,.
\end{align}
$\Gamma = \gamma_5$,
$\gamma_5\gamma_0$ 
for ``good'', $0^+$ diquarks, $\gamma_i$ for
``bad'', $1^+$ diquarks, and $\bf{1}$ and
$\gamma_5\gamma_i$, for the ``not-even-bad'', 
odd-parity $0^-$ and $1^-$ diquarks. We also
measure the correlators of
static-light meson operators $[\bar Q \Gamma q]$.
The static quark $(m_Q\rightarrow\infty)$ acts as a spectator; its mass cancels in mass differences, exposing the diquark spectrum.

We consider diquarks with light-light
$(ud)$, light-strange $(\ell s$, $\ell=u,d)$ and 
strange-strange$^\prime$ $(ss^\prime )$ 
flavors on 5 ensembles with different light-quark,
hence pion, masses. Note in particular that $s^\prime$ denotes a 
hypothetical additional strange valence 
quark. It is introduced to allow a study of 
a good diquark with both quarks having 
the same (strange) quark mass, which the good diquark
flavor antisymmetry makes inaccessible to two
identical $s$ quarks.
Technical aspects of the analysis are summarized in App.~\ref{app:spec}.

\begin{figure}[t!]
\centering
{\includegraphics[width=0.55\columnwidth]{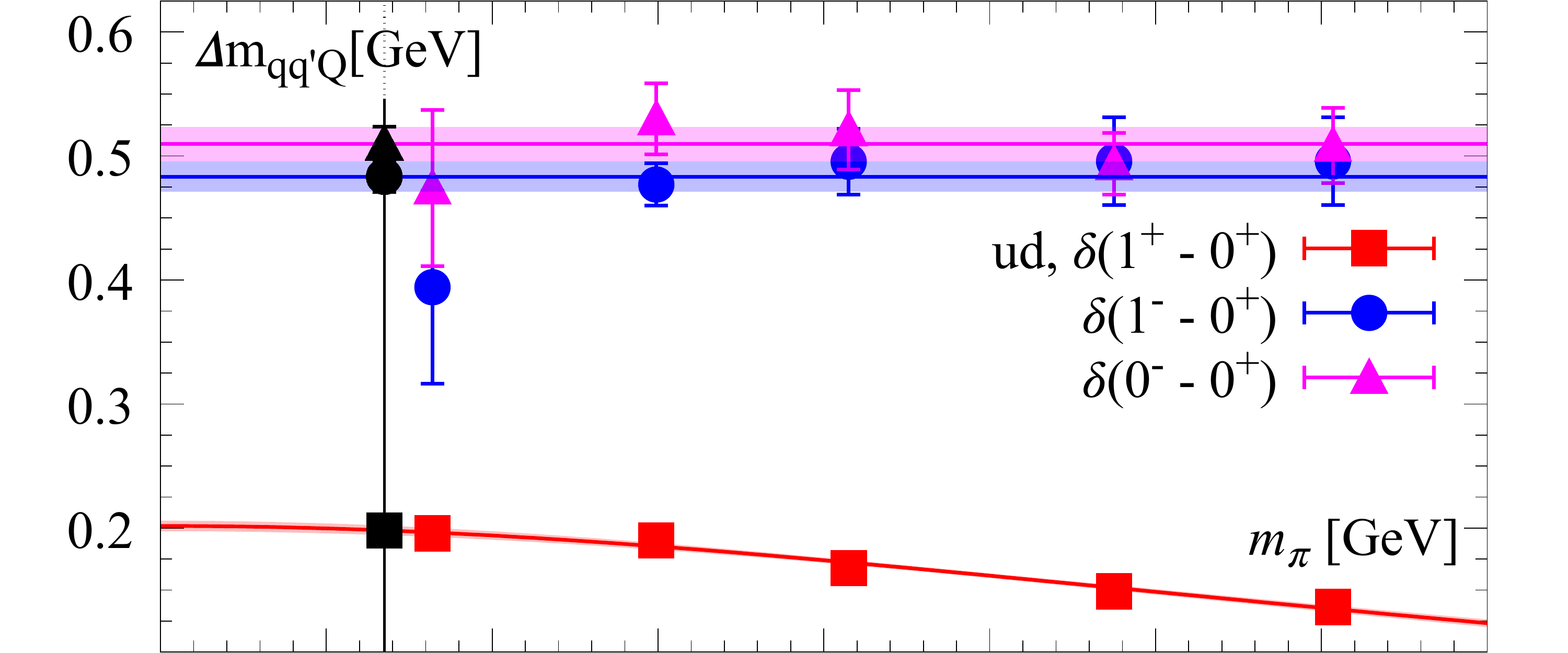}}\\
{\includegraphics[width=0.55\columnwidth]{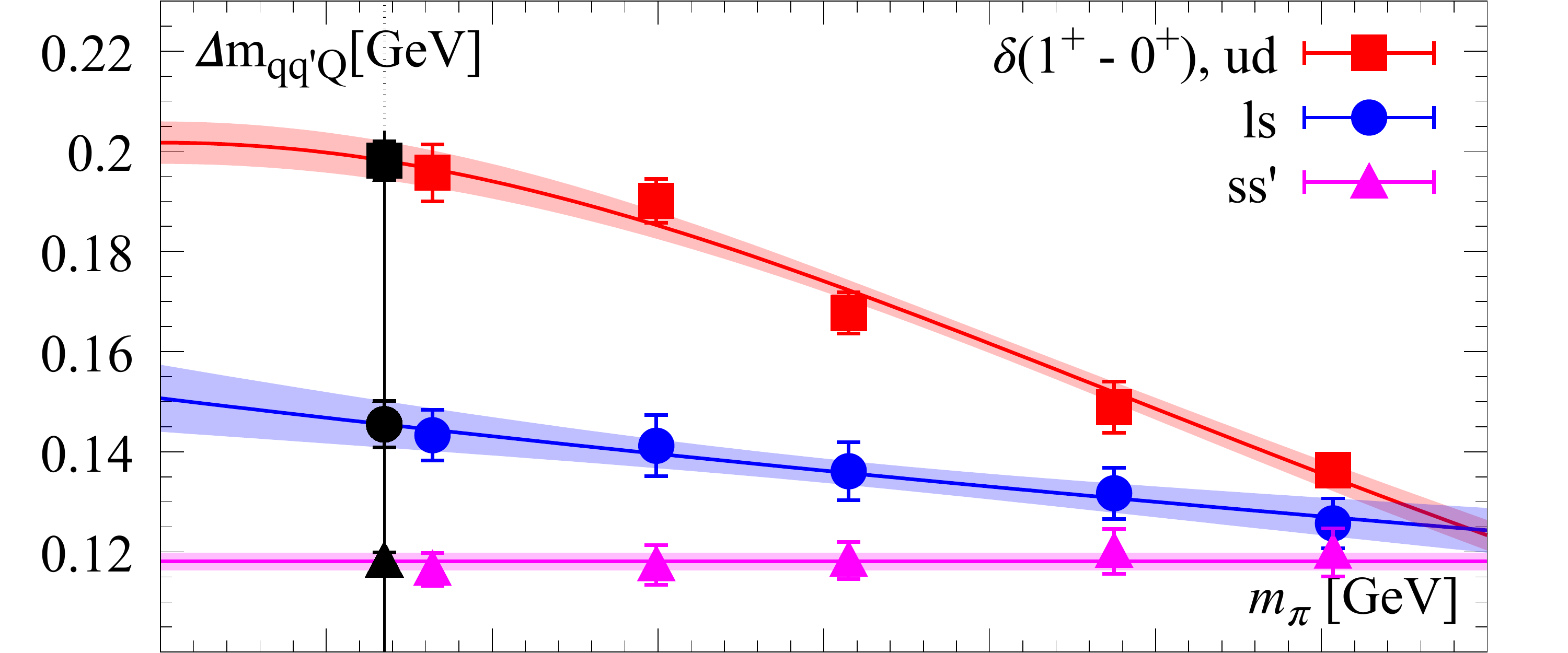}}\\
{\includegraphics[width=0.55\columnwidth]{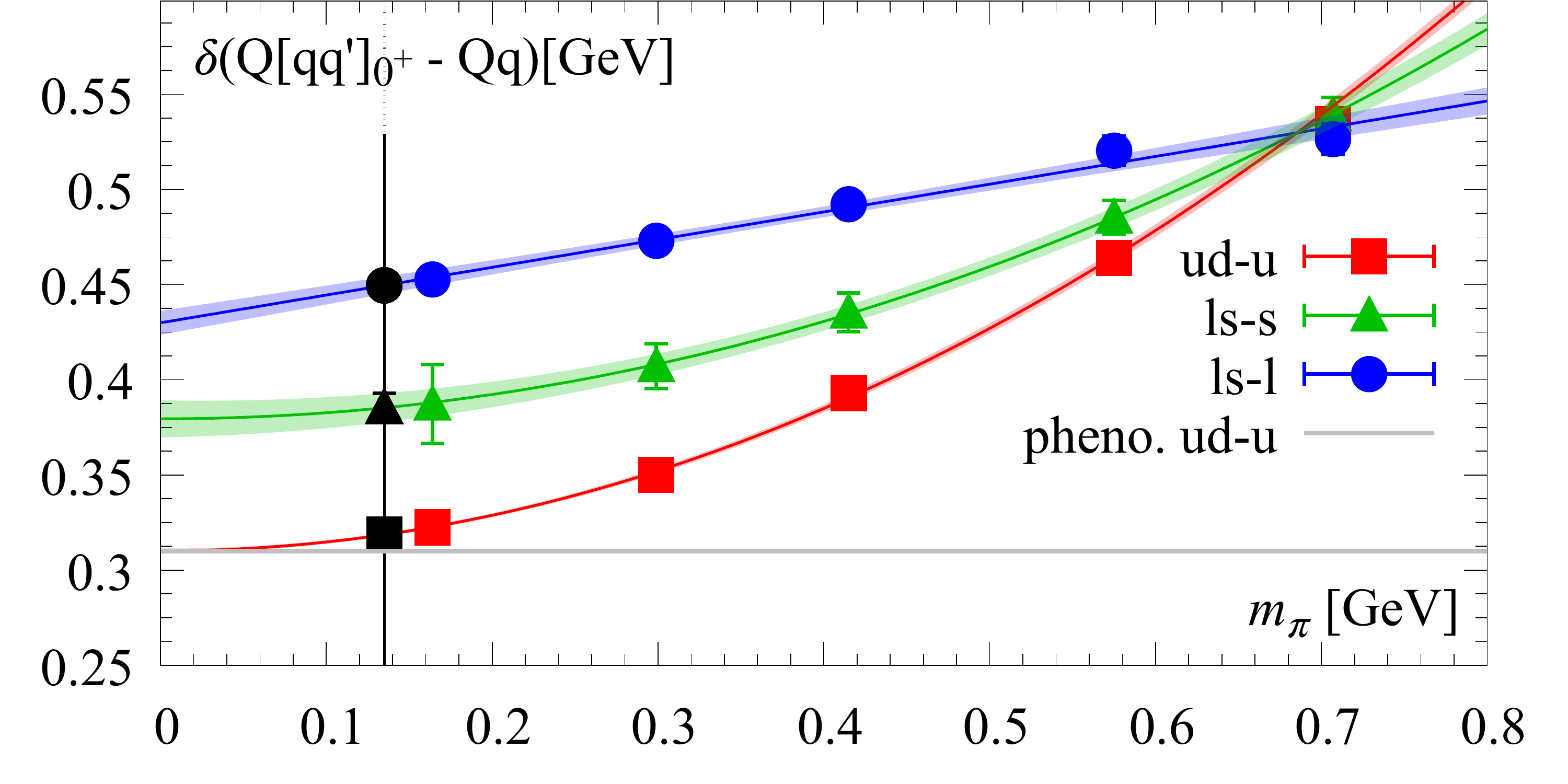}}
\caption{{\it Diquark spectroscopy:
diquark mass differences as a function
of $m_\pi$. Vertical lines identify physical $m_\pi$. The colored bands show the results from fitting to the ans\"atze described in the text.
(Top) The $ud$ differences
$\delta(1^+ - 0^+)$, 
$\delta(1^- - 0^+)$ and $\delta(0^- - 0^+)$.
(Middle) Bad-good $ud$, $\ell s$ and $ss'$ differences, extrapolated to
physical $m_\pi$ using Eq.~(\ref{eq:adiq}).
(Bottom) Analogous $Qqq-\bar{Q}q$ differences, extrapolated 
using Eq.~(\ref{eq:adiq2}). The horizontal line is the phenomenological value 
$\delta(b[u d]_{0^+} - \bar{b} u)=306~\rm{MeV}$.  
}}
\label{fig:spec_panel}
\end{figure}

Figure~\ref{fig:spec_panel} (top panel) 
shows the dependence on $m_\pi$ 
of the $ud$ $0^+$ versus $1^+$, 
$0^-$ and $1^-$ diquark mass differences.
The results provide quantitative 
support for the phenomenological diquark 
approach, which considers only
good $0^+$ diquarks.
Explicitly, the good $0^+$ $ud$ 
diquark lies lowest in the spectrum, 100-200 
MeV below the bad $1^+$ $ud$ diquark. The 
negative-parity $0^-$ and $1^-$ $ud$ diquarks
lie even higher, $\sim 0.5~\rm{GeV}$
above the good diquark and will thus 
play no role in the low-energy physics. The same pattern is observed in the $\ell s$ and $ss^\prime$ sectors.
Figure~\ref{fig:spec_panel} 
(middle panel)
compares the $ud$, $\ell s$ and
$ss^\prime$ $(1^+ - 0^+)$ splittings.
The curves in
Fig.~\ref{fig:spec_panel} are fits 
using Ans\"atze guided by limiting 
cases. Explicitly, the 
$(1^+ - 0^+)$ mass difference goes 
to a constant in the chiral limit 
and decreases as $1/(m_{q_1} m_{q_2})$, 
with $m_\pi \sim (m_{q_1} + m_{q_2})$, 
in the heavy-quark limit. The simplest
interpolation between these
limits is the two-parameter form
\begin{align}
    \delta (1^+-~0^+)_{q_1 q_2} = 
    A/\left[ 1+\big(m_\pi/B\big)^n\right],
    \label{eq:adiq}
\end{align}
with $n=0,1,2$ for 
$q_1 q_2 = ss'$,
$\ell s$ and $ud$, respectively. 
Here, $A$ fixes the chiral 
limit behavior, while $B$ separates
the light- and heavy-quark
regimes. The fits 
clearly describe the data very well. 
A similar Ansatz proposed in \cite{Alexandrou:2005zn},
with $n$ twice as large, produces a much poorer fit.
Note that the $ud$, $\ell s$ and $ss^\prime$ 
curves all intersect at
the flavor-symmetric $n_f=3$ point,
$m_{u,d}\rightarrow m_s$. The 
parameter values and physical-point
mass differences are listed in the 
top half of Table~\ref{tab:Ephys}. 
The latter are in
excellent agreement with phenomenological
expectations~\cite{Jaffe:2004ph} (see
App.~\ref{app:jaffe}).

Further information on the diquark 
spectrum is provided by the mass 
splittings between static $Qqq^\prime$
baryons and the corresponding 
static $\bar{Q}q$,
$\bar{Q}q^\prime$ mesons. 
Results for the $Qud - \bar{Q}u$, 
$Q\ell s - \bar{Q} \ell$ and 
$Q\ell s - \bar{Q}s$
combinations are shown in the bottom panel of Fig.~\ref{fig:spec_panel}, 
together with fits to the Ansatz,
\begin{align}
\delta(Q [q_1 q_2]_{0^+} - \bar{Q} q_2) = 
C~\left[ 1 + (m_\pi/D)^n\right]\, ,
\label{eq:adiq2}
\end{align}
where $C$ fixes the chiral limit value
and $D$ separates the light- and
heavy-$q_1$ quark regimes. In the latter,
the mass splitting must grow linearly 
with the mass $m_{q_1}$, which dictates 
$n=1$ if $q_1$ is a heavy quark,
$n=2$ otherwise. 
The bottom half of Tab.~\ref{tab:Ephys}
lists the fit parameter values and 
resulting extrapolated physical-point 
mass differences.

The excellent agreement with
phenomenological expectations of our results
for all of the $\delta (1^+-0^+)_{ud}$,
$\delta (1^+-0^+)_{us}$, 
$\delta(Q[u d]_{0^+}-\bar{Q}u)$ and
$\delta(Q[u d]_{0^+}-\bar{Q}u)$
splittings, detailed in App.~\ref{app:jaffe},
provides strong evidence that we have
successfully identified the ground-state
heavy baryon signals and that, as
expected, residual discretization effects 
are small. This justifies investigating
the structure of the diquark correlations in 
those baryon ground states using fixed-time 
density-density correlators, described in 
more detail below. Appendix~\ref{app:jaffe} also provides a brief outline of other approaches that
have been used to estimate the good-bad diquark splittings.

An additional interesting relation between the bad-good diquark and $\Delta-N$ splittings is discussed in App.~\ref{app:spec}.

\begin{table}[t!]
    \centering
    \begin{tabular}{cccc}
         \toprule
         All in [GeV]& {$\delta E(m_\pi^{\rm{phys}})~~$} & A& B\\\hline
         $\delta (1^+-~0^+)_{ud\,}$ & 0.198(4) & 0.202(4)& 1.00(5)\\
         $\delta (1^+-~0^+)_{\ell s \,}$ & 0.145(5)& 0.151(7)& 3.7(15)\\
         $\delta (1^+-~0^+)_{ss'}$ & 0.118(2) & 0.118(2)&\\ \hline
         & & C&D\\\hline
         $\delta(Q[u d]_{0^+} - \bar{Q} u)$ & 0.319(1)& 0.310(1)& 0.814(8)\\
         $\delta(Q[\ell s]_{0^+} - \bar{Q} s)$ & 0.385(9)& 0.379(10)& 1.09(6)\\
         $\delta(Q[\ell s]_{0^+} - \bar{Q} \ell)$ & 0.450(6)& 0.430(6)& 2.95(35)\\
         \toprule
    \end{tabular}
    \caption{{\it Fit parameters $A, B, C, D$ and
    physical-point bad-good diquark (top half) 
    and good diquark-quark (bottom half)
mass differences, the errors are statistical only. Here the indices $ud$, $\ell s$, $ss'$ signify the different flavor combinations, with $s'$ denoting a hypothetical additional valence strange quark (see Sec.~\ref{sec:diqspec}). 
For a detailed discussion of the fit ans\"atze, we refer to the text. Further information on phenomenological results can be found in App.~\ref{app:jaffe}.
}}
    \label{tab:Ephys}
\end{table}

\begin{figure}[t!]
\centering
\includegraphics[width=0.55\columnwidth]{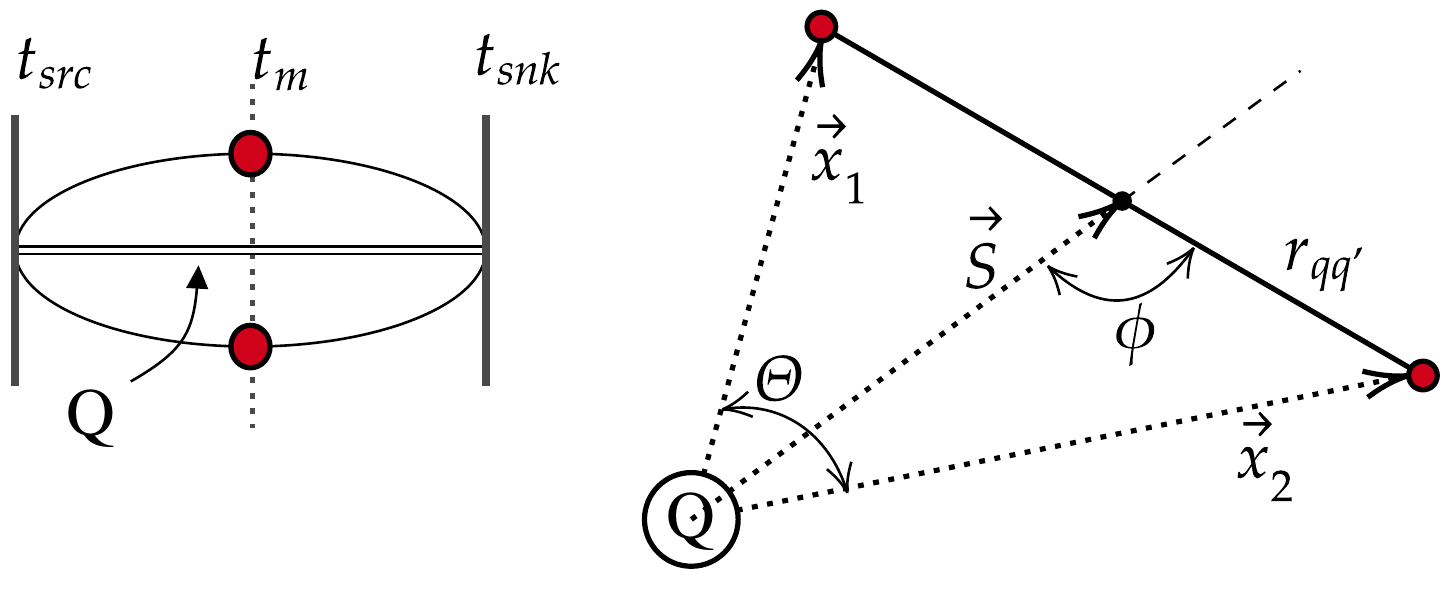}
\caption{{\it Sketch of the 
density correlators. (Left) 2D temporal
view. (Right) Current insertions, spatial view.} }
\label{fig:geom1}
\end{figure}

\section{Diquark structure }
Having successfully identified
the relevant ground-state baryon signals,
we now turn to an investigation
of the light diquark structures in those
states. To do so, we compute the fixed-time 
density-density correlators:
\begin{align}
    C_{\Gamma}^{dd}(\vec x_1, \vec x_2, t)=
    \Big\langle \mathcal{O}_\Gamma(\vec 0,2t)
    \rho(\vec x_1,t)\rho(\vec x_2,t)
    \mathcal{O}_\Gamma^\dagger(\vec 0,0)  \Big\rangle
\end{align}
where $\rho(\vec x,t)=\bar{q}(\vec x,t)\gamma_0 q(\vec x,t)$ 
and $\mathcal{O}_\Gamma$ are the baryon 
operators used before. $\Gamma$
characterizes the diquark channel.
With the static quark at the origin, the
light-quark source and sink points 
are $(\vec 0,t_{src})$ and 
$(\vec 0,t_{snk})$. The currents
are inserted at 
$t_m=( t_{snk}+t_{src} )/2$
to maximize projection onto the
ground state. The relative positions 
of the static source and current
insertions $\vec{x}_1$, $\vec{x}_2$, 
can be characterized by 
$\vec{r}_{qq^\prime}=\vec{x}_2-\vec{x}_1$,
$\vec{S}=(\vec{x}_1+\vec{x}_2)/2$, 
the separation between the static source
and diquark midpoint, and $\phi$, the 
angle between $\vec{r}_{qq^\prime}$
and $\vec{S}$, as shown in 
Fig.~\ref{fig:geom1}. We define 
\begin{align}
    \rho_2(r_{qq^\prime},S,\phi ;\Gamma ) \equiv 
    C^{dd}_\Gamma (\vec x_1, \vec x_2,t_m)\, ,
\end{align}
dropping the label $\Gamma$ when this produces no confusion.

For fixed $S$ and $r_{qq^\prime}$, 
the distance from the static source to the 
closer of the two insertion points is 
minimized (maximized) for 
$\phi =\pi$ ($\pi /2$).
If the proximity 
of a static source disrupts the
diquark correlation in a given 
channel, this disruption will 
thus be largest for $\phi =\pi$ 
and smallest for $\phi =\pi /2$. 
We therefore focus our attention 
on $\rho_2$ for these two 
cases. When $\phi =\pi /2$, 
$\vert \vec{x}_1\vert = 
\vert \vec{x}_2\vert\equiv R$, and
we may instead characterize the relative
positions using $R$ and the angle 
$\Theta$ between $\vec{x}_1$ and 
$\vec{x}_2$. We define 
$\rho_2^\perp (R,\theta )\equiv 
\rho_2(r_{qq^\prime},S,\pi /2)$
and $\rho_2^\parallel(r_{qq^\prime},S)
\equiv \rho_2(r_{qq^\prime},S,\pi )$.
Our calculations
average over all spatial translations. 

\begin{figure}[t!]
\centering
\subfloat{\includegraphics[width=0.5\columnwidth]{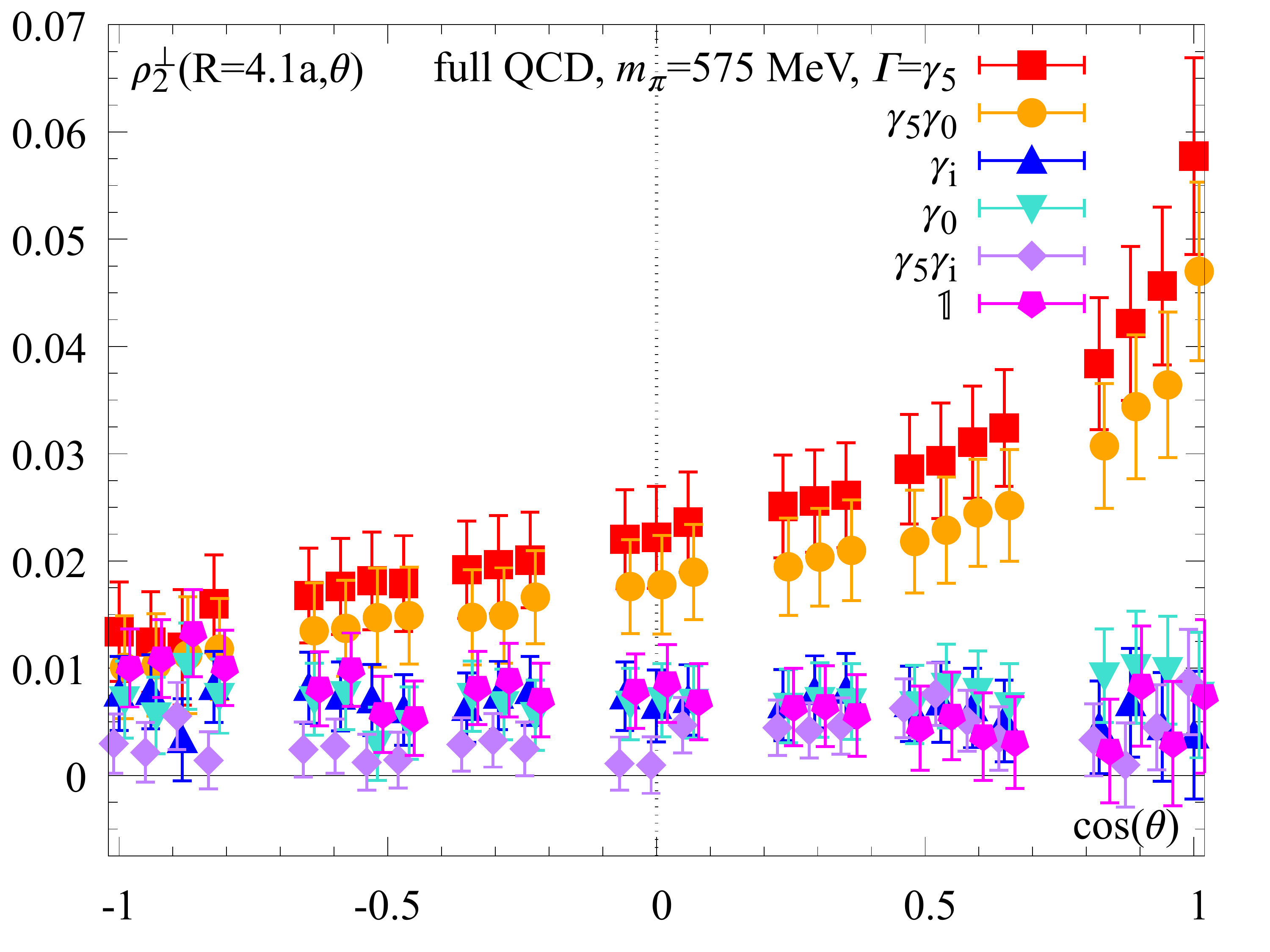}}
\subfloat{\includegraphics[width=0.5\columnwidth]{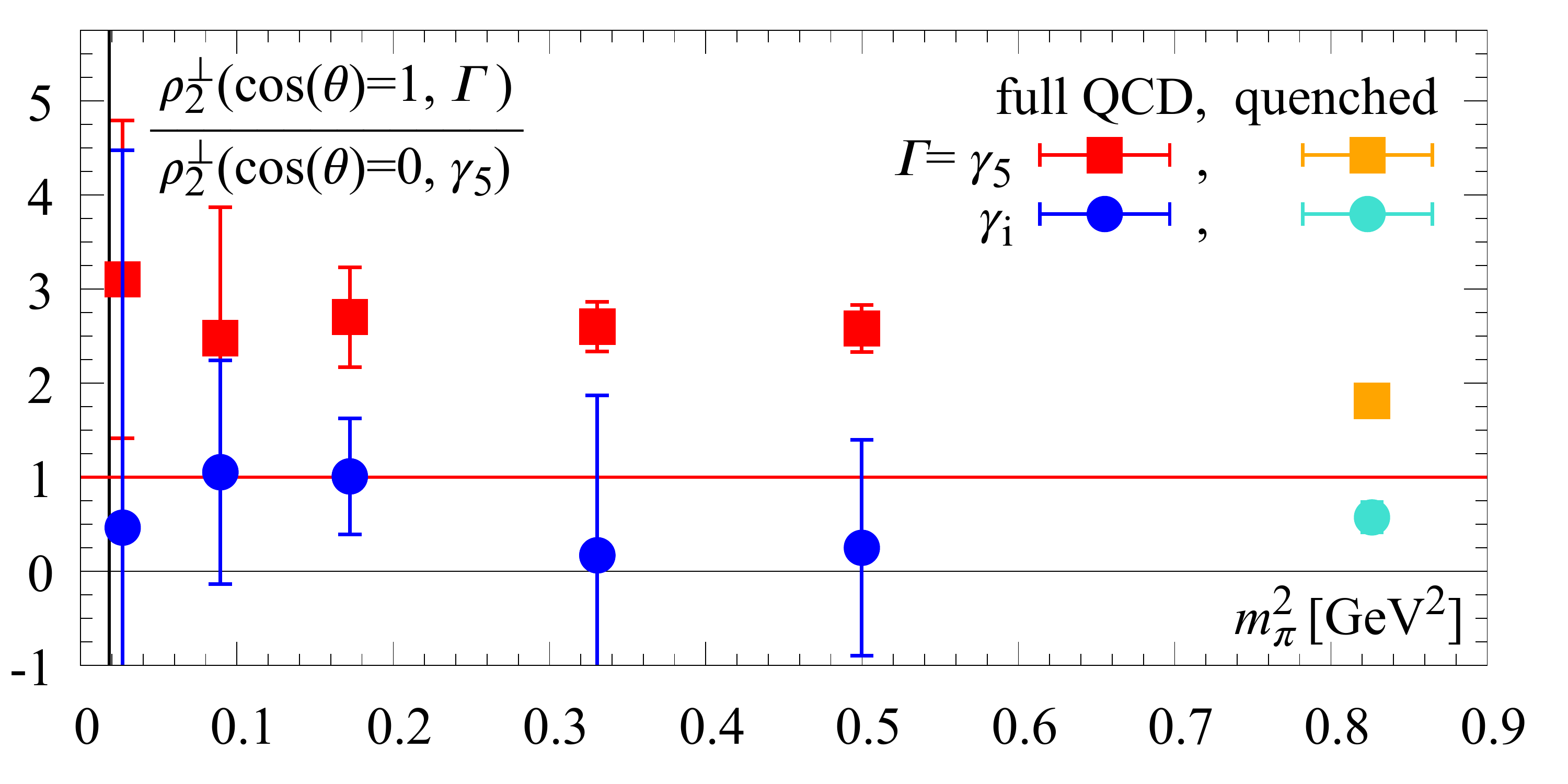}}
\caption{{\it Diquark attractive effect. (Left) 
The density-density correlators
$\rho_2^\perp (R=4.1 a,\Theta ,\Gamma )$ versus 
$cos(\Theta )$ at $m_\pi=575~\rm{MeV}$. 
(Right) The ratio
$\rho_2^\perp (R,\Theta =0,\Gamma )/
\rho_2^\perp(R,\Theta =\pi /2, \Gamma =\gamma_5)$
versus $m_\pi^2$. Values above/below 1 for 
the red/blue points signal
an attraction in the good diquark that is 
absent for the bad diquark. The vertical line
denotes physical $m_\pi$. }}
\label{fig:dens_panel}
\end{figure}

The impact of the light-quark interactions on
the spatial correlation between light quarks
for different $\Gamma$
is displayed in
Fig.~\ref{fig:dens_panel} (left), 
which shows the density-density 
correlations $\rho_2^\perp 
(R,\Theta ,\Gamma )$ as a function of
$\cos(\Theta)$. For illustration, we 
show results for all $\Gamma$
at $R=4.1\, a$ for the ensemble with 
$m_\pi=575~\rm{MeV}$. 
As $\cos(\Theta)$ increases from $-1$ to 
$+1$, $r_{qq^\prime}$ decreases from $2R$ to $0$.
The clear increase in 
$\rho_2$ seen in the good diquark channel is absent in all other channels~\cite{renorm}.
The strengths of the quark-quark attractions in the 
good and bad diquark channels are further 
quantified in 
Fig.~\ref{fig:dens_panel} (right),
which shows the $m_\pi$ dependence of the ratios 
$\rho_2^\perp (R,\Theta =0,\Gamma )/
\rho_2^\perp(R,\Theta =\pi /2, \Gamma 
=\gamma_5)$ for $\Gamma =\gamma_5$ and
$\gamma_i$. The ratio is 
$2$ or more for the good diquark across 
the whole range of $m_\pi$, but consistent 
with $0$ for the bad diquark, with 
no evidence for any $m_\pi$
dependence, apart from a possible low-$m_\pi$
enhancement for the good diquark. The results
confirm a significant attractive
quark-quark spatial correlation in the 
good diquark channel
not present in the bad diquark channel, for all $m_\pi$ studied here.

\begin{figure}[t!]
\centering
\subfloat{\includegraphics[width=0.5\columnwidth]{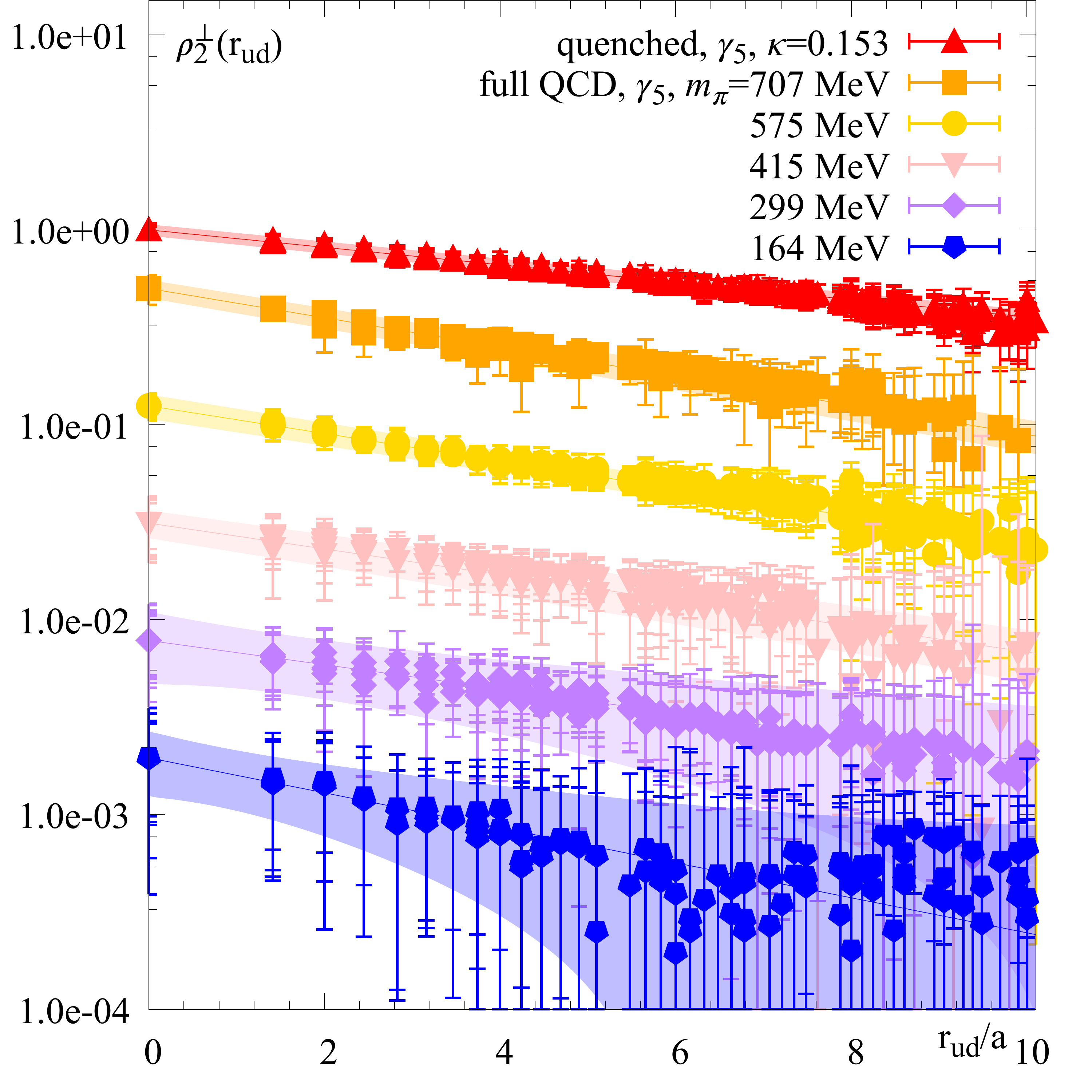}}
\subfloat{\includegraphics[width=0.5\columnwidth]{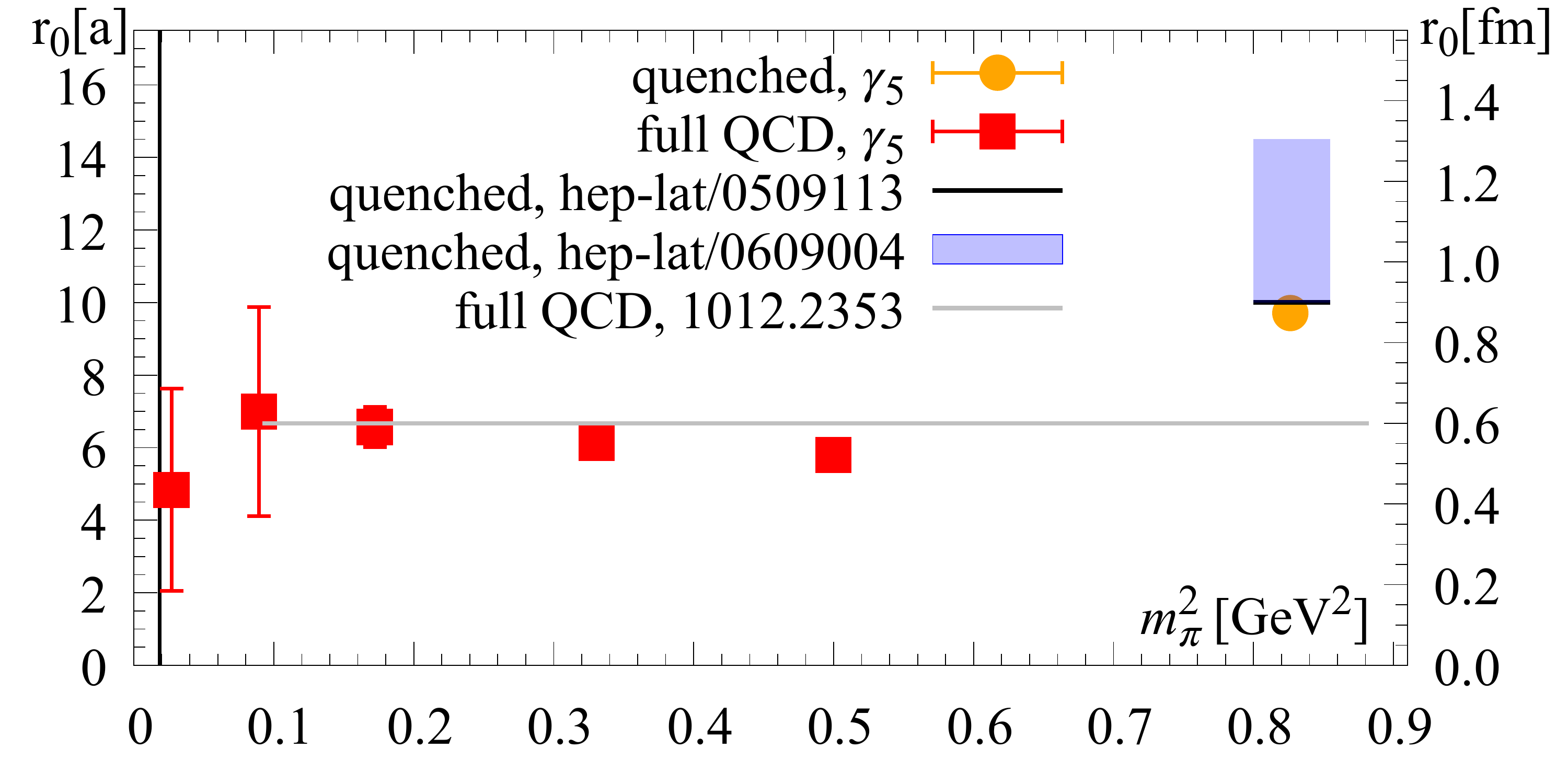}}
\caption{{\it Good diquark size. (Left) 
Exponential decay with $r_{qq^\prime}$ 
of $\rho_2^\perp (R,\Theta )$. 
Each $m_\pi$ has its own color.
Data sets have been normalised at 
$r_{qq^\prime}=0$ and offset vertically. 
Results for all available $R$ are shown 
together in one colored set. Each colored band
comes from the combined fit used to determine the diquark size $r_0(m_\pi^2)$. (Right) 
Resulting good diquark size $r_0$ versus
$m_\pi^2$, compared to results of other lattice
studies in the literature. The vertical line denotes physical
$m_\pi$. } }
\label{fig:rad_panel}
\end{figure}

With a significant attractive good 
diquark spatial correlation 
established, we can 
refine our picture of the good diquark by
studying its size and shape. We consider 
first the case $\phi =\pi /2$. At fixed $R$, 
$\rho_2^\perp (R,\Theta ,\Gamma = \gamma_5)$ 
depends only on $\Theta $ or, equivalently, 
$r_{qq^\prime}=R\sqrt{2(1-\cos(\Theta ))}$. 
We find this dependence
well represented by 
an exponential form, $\rho_2^\perp
(R,r_{qq^\prime})\sim 
\exp(-r_{qq^\prime}/r_0)$,
for each value of $R$. As $R$ decreases 
and the diquark moves closer to the static 
quark, one might 
expect the latter
to distort 
such diquark correlations and 
cause $r_0$ to vary. We see no evidence 
for such a variation, so long as 
$R > r_{qq^\prime}$, and thus, in the left 
panel of Fig.~\ref{fig:rad_panel}, display
results for all $R$ together, for each 
$m_\pi$. We take $r_0$ as our definition 
of the good diquark size and fix its 
value from a combined fit to data 
for all such $R$. The 
resulting $r_0(m_\pi^2)$ are displayed,
and compared to those obtained in
Refs.~\cite{Alexandrou:2005zn,Green:2010vc},
in the right panel of 
Fig.~\ref{fig:rad_panel}.
Recall that the parameters of 
our quenched ensemble match exactly 
those of
\cite{Alexandrou:2005zn}. Our
results are in very good agreement with 
\cite{Alexandrou:2005zn}, and with both 
the quenched and dynamical results of
\cite{Green:2010vc}.

Increasing $m_{q_{1,2}}$
should, on its own, produce a more 
compact object. The accompanying decrease 
in good diquark attraction seen in
Tab.~\ref{tab:Ephys} will, however,
work in the opposite direction.
We see some evidence that the former 
effect dominates for larger $m_\pi$ 
(above $\sim 400$ MeV) though, 
in the limit of infinitely massive sea 
quarks ({\it i.e.} the quenched case) 
the diquarks are definitely larger.
As Fig.~\ref{fig:rad_panel} (right) indicates, in full 
QCD, over a range of $m_\pi$, $r_0$ is of the
order of $0.6~\rm{fm}$, a size similar to that
of static-light mesons when measured the same 
way~\cite{Blossier:2016vgh}. 
%
%
This result is also in good agreement with that of a phenomenological, relativistic 
quark-diquark model study of nucleon form factors~\cite{DeSanctis:2011zz}, which 
found a fitted $0^+$ diquark form factor corresponding to an rms diquark size of 
$\sim 0.54$ fm.
It should be noted that the determined diquark size does not affect the spectroscopy of models that include diquarks as effective degrees of freedom. 
Our results clearly support diquark modelling of the baryon
structure which allows for the possibility of a non-zero diquark size.

\begin{figure}[t!]
\centering
\subfloat{\includegraphics[width=0.55\columnwidth]{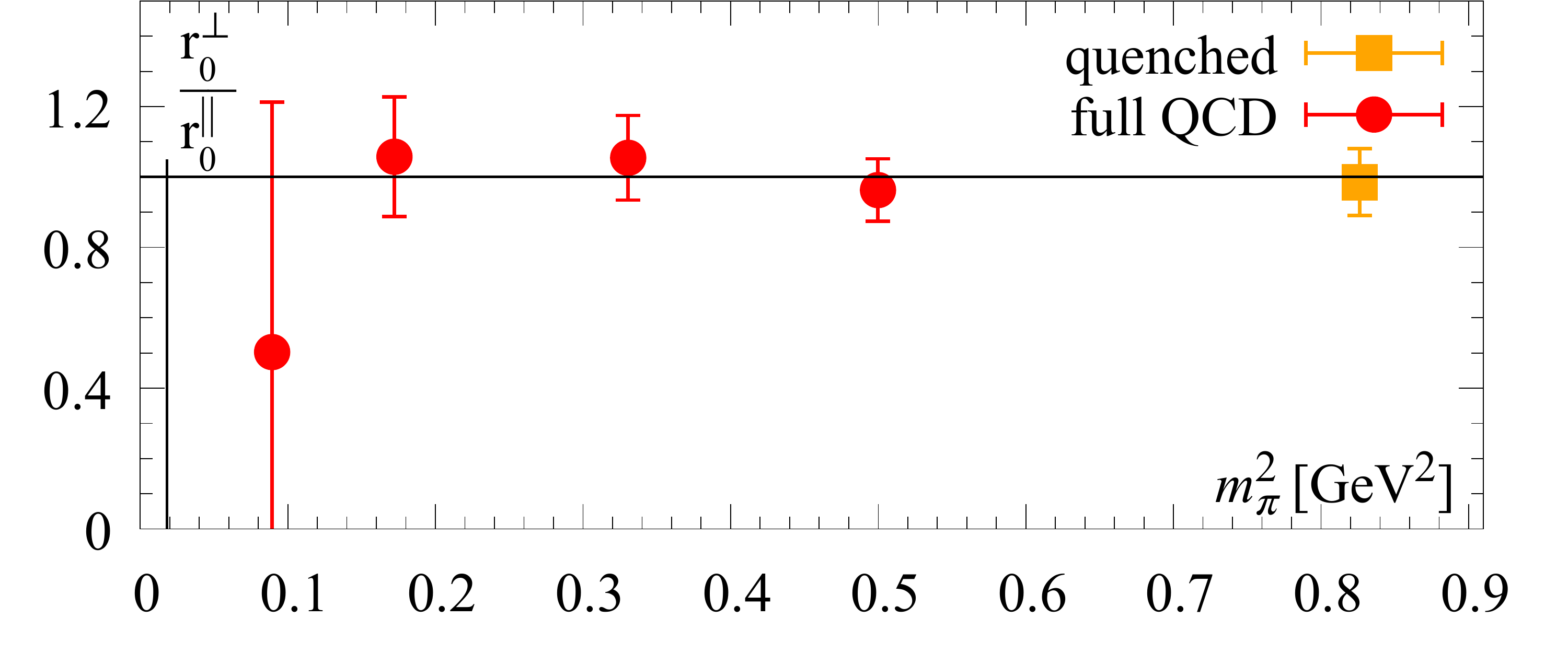}}
\caption{
{\it Good diquark shape. The ratio 
$r_0^\perp/r_0^\parallel$ as
a function of $m_\pi$. 
The vertical line denotes physical 
$m_\pi$.}
}
\label{fig:obl_panel}
\end{figure}

Finally, we can learn about the good diquark 
shape, by comparing the density-density correlation 
falloff for the relative radial ($\phi =\pi$)
and tangential ($\phi =\pi /2$) orientations 
of $\vec{x}_2-\vec{x}_1$ and $\vec{S}$
(sketched in Fig.~\ref{fig:geom23} of 
App.~\ref{app:struct}).
We define
separate {\it radial} (${\parallel}$) 
and {\it tangential} ($\perp$) size parameters,
$r_0^\parallel$ and $r_0^\perp$, from 
exponential fits to the data for 
$\rho_2^\perp (R,\Theta )$ and 
$\rho_2^\parallel\, (r_{qq^\prime},S)$, 
detailed in App.~\ref{app:struct} and shown
in the left and right
panels of Fig.~\ref{fig:obl_panel_supp}.

The ratio $r_0^\perp/r_0^\parallel$
provides a measure of whether 
the diquarks are prolate, oblate, or 
neither. The results are shown in 
Fig.~\ref{fig:obl_panel}.
We find $r_0^\perp/r_0^\parallel(m_\pi^2)
\simeq 1$ within errors for all $m_\pi$,
indicating that the diquarks have a
near-spherical shape. This is consistent 
with the scalar, $J=0$ nature of the 
good diquark, though the presence of 
the static quark could, 
in principle, have induced a diquark 
polarization. There appears no need 
to include a dipole term in diquark models.

\section{Summary and conclusions}
Using a gauge-invariant setup, we have studied 
the masses and shapes of diquarks carrying 
different quantum numbers. Our study is the 
first to consider $n_f = 2+1$ flavors of 
dynamical quarks with a range of $u,d$ 
masses corresponding to $m_\pi$ as low as 
$164~\rm{MeV}$. This allows for a small,
controlled extrapolation to physical 
$m_\pi \approx 135~\rm{MeV}$. The resulting
diquark mass differences 
presented in 
Fig.~\ref{fig:spec_panel} and
Tab.~\ref{tab:Ephys} confirm the 
special status of the ``good'' diquark channel, 
which shows an attraction of $198(4)~\rm{MeV}$
over the ``bad'' channel, more over the others. 
A simple interpolation Ansatz
Eq.~(\ref{eq:adiq}) accurately describes 
how this attraction varies with $m_\pi$, 
and with the diquark flavor composition.
Extrapolation of our results to the 
continuum limit is still required, 
but this has been 
found to amount to a small correction, at 
the percent level, in other hadronic mass
measurements on the same gauge 
configurations~\cite{Padmanath:2019ybu,Namekawa:2013vu,Alexandrou:2017xwd,Hudspith:2017bbh}.
We have also measured the mass difference
between a good diquark and an [anti]quark, 
as per Tab.~\ref{tab:Ephys}. 

We have also
shown that the $q-q$ attraction 
responsible for the bad-good diquark 
mass differences induces a compact 
spatial correlation, present in the ``good''
diquark channel only. The associated ``good''
diquark size, extracted from the spatial 
decay rate of quark density-density 
correlations, is ${\cal O}(0.6)$~fm, similar 
to that of ordinary mesons and baryons 
\cite{Blossier:2016vgh}, and varies 
little with light-quark mass. 

Finally, we have tried to refine the diquark 
picture further, by studying the 
shape
of quark density-density
correlations in a good diquark, in the
background of a heavy, static quark. It turns 
out that good diquarks are nearly spherical,
with no signal within errors of a departure 
from this simplest shape.

The information obtained above may prove useful, both in identifying
channels favorable to the existence of low-lying tetraquark or pentaquark
states, and in obtaining rough estimates of their expected masses. Such
qualitative guidance has, in fact, already been exploited in identifying
double-open-heavy-flavor, $SU(3)_F$ flavor $\bar{3}_F$, $J^P=1^+$
$\bar{Q}\bar{Q}^\prime qq^\prime$ channels as favorable to the existence
of exotic tetraquark states. In such channels, a localized four-quark
configuration benefits from the attractive good-light-diquark and
color $3_c$ heavy-antidiquark Coulomb interactions, neither of which is
accessible for two well-separated heavy-light mesons. This observation
motivated both phenomenological and lattice explorations of potential
binding in such doubly heavy tetraquark
channels\footnote{See Refs.~\cite{Karliner:2017qjm,Eichten:2017ffp,Czarnecki:2017vco,Wagner:2011ev,Brown:2012tm,Bicudo:2015kna,Bicudo:2012qt,Bicudo:2015vta,Bicudo:2016ooe,Francis:2016hui,Francis:2018jyb,Junnarkar:2018twb,Leskovec:2019ioa,Hudspith:2020tdf,Mohanta:2020eed,Bicudo:2021qxj} and earlier references therein.}, and experimental searches
for bound doubly heavy tetraquark states, the latter culminating in the
LHCb discovery of the exotic doubly charmed $T_{cc}$ tetraquark 
state~\cite{LHCb:2021vvq,LHCb:2021auc}. Multiple recent lattice studies using interpolating
operators designed to access the expected good-light-diquark configuration
now also provide clear evidence for the existence of a $J^P=1^+$, $SU(3)_F$
$\bar{3}_F$ multiplet of doubly bottom strong-interaction-stable tetraquark states~\cite{Francis:2016hui,Bicudo:2015kna,Bicudo:2012qt,Bicudo:2016ooe,Junnarkar:2018twb,Leskovec:2019ioa,Hudspith:2020tdf,Mohanta:2020eed,Bicudo:2021qxj}. An analogous
qualitative diquark-based argument identifies the singly-heavy $J^P=1/2^+$,
$I=1/2$, $\bar{Q}sudd$ channel as one potentially favorable to the existence
of an exotic pentaquark resonance. Explicitly, while at most one good light
diquark can exist in a state consisting of a well-separated heavy-light meson
and light-quark baryon, a localized singly heavy five-quark state can contain
two good light diquarks, one non-strange and one strange.{\footnote{In such
a singly heavy pentaquark channel, the four light quarks can be organized
into two good light diquark pairs only if the four-quark spin and color are
$0$ and $3_c$, respectively. To satisfy Pauli statistics, a low-lying state
with no internal spatial excitation must then have four-quark flavor $3_F$,
and hence contain at least one $u$, one $d$ and one $s$ quark.}} The
possibility that the short-distance part of the associated singly heavy
meson-light baryon system might have an attractive component resulting
from this localized "extra-good-light-diquark" configuration motivates
further study of this channel. Outstanding issues still to be investigated
are potential distortions of the good diquark correlation caused by the
presence of additional light quarks and/or the impact of Pauli blocking
in channels, like this, where more than one good light diquark may be
present. These are questions that might be amenable to investigation using
microscopic models which survive the tests of predicting very shallow
binding in the $T_{cc}$ channel and binding energies compatible with now
well-established lattice results for the non-strange and strange doubly
bottom $J^P=1^+$, $\bar{3}_F$ channels. Bound or resonant singly-heavy
$J^P=1/2^+$, $I=1/2$, $\bar{Q}sudd$ pentaquark states, if they exist,
would have four open flavors and hence, like the doubly heavy tetraquarks,
be manifestly exotic.

Our three sets of results 
-- bad-good diquark 
mass difference, good diquark size and shape --
paint a diquark picture entirely consistent 
with that used in diquark models
and provide clear, quantitative
support for the good diquark picture. Diquark 
models are playing an important role in
explorations of possible multi-quark exotics, 
especially in channels too complex to permit a 
complete theoretical analysis. In such channels,
various light quark pairings into compact good 
diquark composites are likely to
occur, especially when heavy $c, b$ quark
sources are present (see {\it e.g.}
\cite{Karliner:2017qjm,Eichten:2017ffp,Czarnecki:2017vco} for phenomenological and \cite{Wagner:2011ev,Brown:2012tm,Bicudo:2015kna,Bicudo:2012qt,Bicudo:2015vta,Bicudo:2016ooe,Francis:2016hui,Francis:2018jyb,Junnarkar:2018twb,Leskovec:2019ioa,Hudspith:2020tdf,Mohanta:2020eed,Bicudo:2021qxj} for lattice studies).
Which pairing is energetically favored 
depends sensitively on the numerical 
values of the parameters of the diquark model. 
Our study may sharpen these values, and 
help improve the reliability
of such diquark analyses.

\vspace{1ex}
\section*{Acknowledgements}
Calculations were performed on the HPC clusters HPC-QCD@CERN 
and Niagara@SciNet supported by Compute Canada. AF thanks M. Bruno, B.~Colquhoun, P.~Fritzsch, J.~Green and M.~Hansen for discussions. PdF thanks R. Fukuda for his help with an earlier, not completed version of 
this project \cite{Fukuda:2017mmh}, and K.
Fukushima for discussions.
RL and KM acknowledge the support
of grants from the Natural Sciences and 
Engineering Research Council of Canada. Together we thank R.~J.~Hudspith for reviewing an initial version of this manuscript.


\appendix

\section{Phenomenological Expectations}
\label{app:jaffe}

Ref.~\cite{Jaffe:2004ph} discussed in detail how to obtain phenomenological estimates
for the static-limit values of the bad-good diquark and diquark-antiquark
mass differences using combinations of single-charm and single-bottom meson
and baryon masses chosen so $O(1/m_Q)$ contributions cancel, bringing
the results closer to the static limit. Comparing the estimates for a given
splitting obtained using charm input to that obtained using bottom input
provides an assessment of how close to the static limit the bottom-based
estimate is likely to be. This data-based approach is obviously very
closely related to the gauge-invariant, static limit approach used to
obtain our lattice results above. In this appendix we provide an update
of the numerical analysis of Ref.~\cite{Jaffe:2004ph}. We also remind the reader of a
number of other, generally more model-dependent approaches, that have
been used to obtain estimates of the good-bad diquark splittings.


Ref.~\cite{Jaffe:2004ph} gives expressions
for the combinations needed to provide
phenomenological estimates for four of the
splittings we have measured. Explicitly,
the combination
\begin{equation}
    \frac{1}{3}\left( 2M(\Sigma^{*}_{Q})+M(\Sigma_{Q})\right) -M(\Lambda_{Q})
\label{udbadgood}
\end{equation}
provides an estimate for 
$\delta (1^+-0^+)_{ud}$, the combination
\begin{equation}
    \frac{2}{3}\left( M(\Xi^{*}_{Q})+M(\Sigma_{Q})
+M(\Omega_{Q} ) \right) -M(\Xi_{Q})-M(\Xi'_{Q})
\label{usbadgood}
\end{equation}
an estimate for $\delta (1^+-0^+)_{us}$,
the combination
\begin{equation}
    M(\Lambda_{Q})-\frac{1}{4}\left(
    M(P_{Qu})+3M(V_{Qu})\right)\, ,
\label{udgoodubar}
\end{equation}
with $P_{Qu}$ and $V_{Qu}$ the ground-state, 
heavy-light pseudoscalar and vector
mesons,
an estimate for $\delta(Q[ud]_{0^+}-\bar{Q}u)$,
and the combination
\begin{eqnarray}
&&M(\Xi_{Q})+M(\Xi'_{Q})
-\frac{1}{2}(M(\Sigma_{Q})+M(\Omega_{Q}))
\nonumber\\
&&\qquad\qquad
-\frac{1}{4}(M(P_{Qs})+3M(V_{Qs}))\, ,
\label{usgoodsbar}
\end{eqnarray}
with $P_{Qs}$ and $V_{Qs}$ the ground-state,
heavy-strange pseudoscalar and vector mesons, 
an estimate for $\delta(Q[us]_{0^+}-\bar{Q}s)$.
For a given static-limit splitting, the 
most accurate estimate should be that obtained 
using bottom hadron input, while the 
difference between the charm- and bottom-based
estimates should provide a conservative 
assessment of the deviation of the bottom-based
estimate from the actual static-limit value. 
At the time Ref.~\cite{Jaffe:2004ph} was 
written, information on bottom hadron masses
was limited, and only one of these four
splittings, $\delta(Q[u d]_{0^+}-\bar{Q}u)$,
could be estimated with both charm and bottom
input. The agreement between the two was
excellent. It is now possible to estimate
all four splittings using both charm and
bottom input. Using PDG 2021 input~\cite{Zyla:2020zbs}, we find, for 
$\delta (1^+-0^+)_{ud}$, $210$ MeV using
charm input and $206$ MeV using bottom
input; for $\delta (1^+-0^+)_{us}$, $148$ MeV
using charm input and $145$ MeV using bottom
input; for $\delta(Q[u d]_{0^+}-\bar{Q}u)$,
$313$ MeV using charm input and $306$ MeV
using bottom input; and, for 
$\delta(Q[u s]_{0^+}-\bar{Q}s)$ $398$ MeV
using charm input and $397$ MeV using bottom
input. It follows that the bottom-based 
phenomenological estimates for the static-limit
splittings should be reliable to 
$O(7)\ {\rm MeV}$ or better. 
Moreover, these estimates agree well
with the results shown in
Tab.~\ref{tab:Ephys}.

A number of other approaches have also been used to estimate the
$ud$ and $\ell s$ good-bad diquark splittings.

In the Dyson-Schwinger equation (DSE) approach, earlier analyses employing
the rainbow-ladder approximation obtained $\delta (1^+ - 0^+)_{ud}=212$ MeV
and $\delta (1^+ - 0^+)_{us}=168$ MeV~\cite{Burden:1996nh},
$\delta (1^+ - 0^+)_{ud}=202$ MeV~\cite{Maris:2005zy},
$\delta (1^+ - 0^+)_{ud}=270(30)$ MeV~\cite{Eichmann:2008ef},
$\delta (1^+ - 0^+)_{ud}=280$ MeV~\cite{Roberts:2011cf}.
A more recent analysis, Ref.~\cite{Eichmann:2016hgl}, reports a result
$\delta (1^+ - 0^+)_{ud}=190(20)$ MeV, in good agreement with both our
lattice determination and the updated version of the phenomenological
estimate of Ref.~\cite{Jaffe:2004ph}.

The good-bad diquark mass splittings have also been obtained from
diquark-quark model analyses of the non-strange and strange baryon
spectrum in which the diquark mass enters as a free parameter of the
model and is obtained as part of the fit to the spectrum. An iterative,
phenomenological version of this approach~\cite{Lichtenberg:1996fi}
produced the results $\delta (1^+ - 0^+)_{ud}=205$ MeV and
$\delta (1^+ - 0^+)_{us}=140$ MeV, while a more microscopic,
relativistic model, which did not, however, allow for the possibility
of mixing between quark-scalar-diquark and quark-axial-vector-diquark
configurations, obtained a larger result, $\delta (1^+ - 0^+)_{ud}=350$ MeV,
for the $ud$ diquark splitting~\cite{Ferretti:2011zz}. A modified version
of this model, which significantly improves the quality of the model fit
to known $3^*$ and $4^*$ baryon resonances, obtained by adding a term to
the effective interaction that allows such mixing to occur, in contrast,
produces a result $\delta (1^+ - 0^+)_{ud}=210$ MeV in good agreement
with both our lattice determination and the updated version of the
phenomenological estimate of Ref.~\cite{Jaffe:2004ph}.

An alternate implementation of the microscopic quark-diquark model
approach first fits the parameters of a model two-body quark-antiquark
effective interaction with one-gluon-exchange color dependence,
$F_q\cdot F_{\bar{q}}$, to the meson spectrum, then uses this interaction,
with $F_q\cdot F_{\bar{q}}$ replaced by the corresponding quark-quark
one-gluon-exchange factor, $F_q\cdot F_q$, to determine the nominal
$\bar{3}_c$, $J^P=0^+$ and $1^+$ masses, and hence the $1^+$-$0^+$
splittings. The meson sector model used in Ref.~\cite{Ebert:2011kk}
(which has a two-body confinement interaction involving a linear
combination of scalar and vector structures) produces the results
$\delta (1^+ - 0^+)_{ud}=199$ MeV and $\delta (1^+ - 0^+)_{us}=121$
MeV~\cite{Ebert:2011kk}, while the Godfrey-Isgur model~\cite{Godfrey:1985xj}
used in Ref.~\cite{Ferretti:2019zyh} (with its purely scalar two-body
confinement form) produces the results $\delta (1^+ - 0^+)_{ud}=149$ MeV
and $\delta (1^+ - 0^+)_{us}=106$ MeV.

In view of the good agreement between the updated versions of
the charm- and bottom-based estimates of Ref.~\cite{Jaffe:2004ph}, we consider
the bottom-based results to represent the best phenomenological
estimates of the gauge-invariant static limit splittings we measure
on the lattice. The other approaches, which are more model-dependent,
but have the advantage of being applicable to non-strange and strange
baryon sector, produce results in reasonable to good agreement
with the heavy-quark based phenomenological estimates for the
good-bad diquark splittings, with more recent versions of the
analyses typically producing improved agreement. It is, of course,
possible that the additional light quarks present in non-strange
and strange baryons might affect the structure of the good light
diquark correlation in those systems, causing it to differ
from that found in singly heavy baryon systems. The agreement
of the results for the diquark splittings obtained from (albeit
model-dependent) analyses of the light baryon sector with the
heavy-quark-based phenomenological estimates is thus of interest
since it supports the picture in which the same good light diquark
correlations serve as useful effective degrees of the freedom in
light and heavy baryon sectors.

\section{Lattice ensembles and propagators}
\label{app:latt}

For the numerical studies, we re-use the set 
of propagators from
\cite{Francis:2016hui,Francis:2018jyb} 
determined on the publicly available $n_f=2+1$ 
flavor full QCD gauge ensembles 
provided by the PACS-CS'09 collaboration \cite{Aoki:2008sm,Namekawa:2013vu} 
via the JLDG repository \cite{JLDG}. The quoted values of $m_\pi$ and 
lattice spacing originate from our own previous re-determination \cite{Hudspith:2017bbh,Francis:2016nmj}. These gauge configurations have been 
used extensively within the lattice community. A known caveat is the
slight mistuning of the strange sea quark mass \cite{Aoki:2008sm}. 
The value of the hopping parameter which produces the physical strange quark
mass is, however, known, and we set the strange valence quark mass 
to this value, thus introducing a tiny amount of partial quenching.

To connect with previous studies in the quenched setup discussed in 
\cite{Alexandrou:2006cq}, and in more detail in \cite{Alexandrou:2005zn}, 
we generated a new ensemble with the same lattice parameters, in particular
with coupling $\beta=6.0$ and hopping parameter $\kappa=0.153$ for 
propagator inversions. This corresponds to a valence pion mass 
$m_\pi^v=909\,\rm{MeV}$. 

All propagators were computed using the deflated SAP-GCR solver 
\cite{CLScode} and have Coulomb gauge-fixed wall sources, where 
the gauge was fixed using the FACG-algorithm in the implementation 
of \cite{Hudspith:2014oja,GLUcode}.

We can re-use the propagators without further inversions since the 
gauge-fixed wall sources enable the contraction of the density 
correlators without an additional sequential source propagator. 
Choosing $( t_{snk}-t_{src} )=16$ enables us to perform multiple measurements on each configuration. Setting $t_m$ midway between source and sink minimizes excited-state contamination. 

For the static quark we compute propagators on the fly via 
($t_2>t_1$)\cite{DellaMorte:2005nwx}:
\begin{align}
    S(\bm{x},t_2,\bm{x},t_1)=\Big(\frac{1+\gamma_0}{2}\Big)
    \Big[ \prod_{t=t_1}^{t_2-a} U_0(\bm{x},t)\Big]
\end{align}
where we dropped the exponential prefactor, since it amounts to a 
constant shift in the masses that either drops out in the difference or 
is irrelevant to the results. To reduce statistical fluctuations, 
the gauge links are smeared using HYP smearing 
(type 1 \cite{Donnellan:2010mx,Hasenfratz:2001hp}) in all 4 dimensions, 
which introduces some non-locality in time. In all cases we made sure 
that the propagation time $t$ is large enough and the number of 
smearing steps small enough to ensure negligible effects aside 
from the boosted signal. Furthermore we considered several 
smearing setups and smearing radii. We observed 
comparable results and selected the one giving 
the best signal-to-noise properties. 
Our lattice parameters are listed in Tab.~\ref{tab:latpar}.

\begin{table}[t!]
    \centering
    \begin{tabular}{cccccccc}
         \toprule
         Label & $L\times T$ & $a^{-1}[\rm{GeV}]$ & $m_\pi[\rm{MeV}]$ 
& $n_{cfg}$ & $n_{meas}^{spec}$ & $n_{meas}^{struct}$ \\\hline
         Q & $32\times 64$ & $2.15$ & 909 & 374 & 374 & 374 \\\hline
         E1 & $32\times 64$ & $2.194$ & 707 & 399 & 1596 & 798\\
         E2 & " & " & 575 & 400 & 1600 & 800\\
         E4 & " & " & 415 & 400 & 3200 & 2800\\
         E5 & " & " & 299 & 800 & 6400 & 5600\\
         E6 & " & " & 164 & 198 & 6336 & 2574\\
         \toprule
    \end{tabular}
    \caption{{\it Parameters of the lattice calculation. $n_{meas}^{spec}$ and 
$n_{meas}^{struct}$ indicate how many measurements in total were made of 
the baryon/meson correlators for the spectroscopy study, and of the 
density-density correlators for the structure analysis, respectively. For
the latter analysis, the sink-source time propagation is set to $( t_{snk}-t_{src} )=16$, 
with the currents inserted at $t_m=8$, see also the
sketch in Fig.~\ref{fig:geom1} (left). }}
    \label{tab:latpar}
\end{table}

\section{Lattice spectroscopy analysis details}
\label{app:spec}

\begin{figure}[t!]
\centering
\includegraphics[width=0.55\columnwidth]{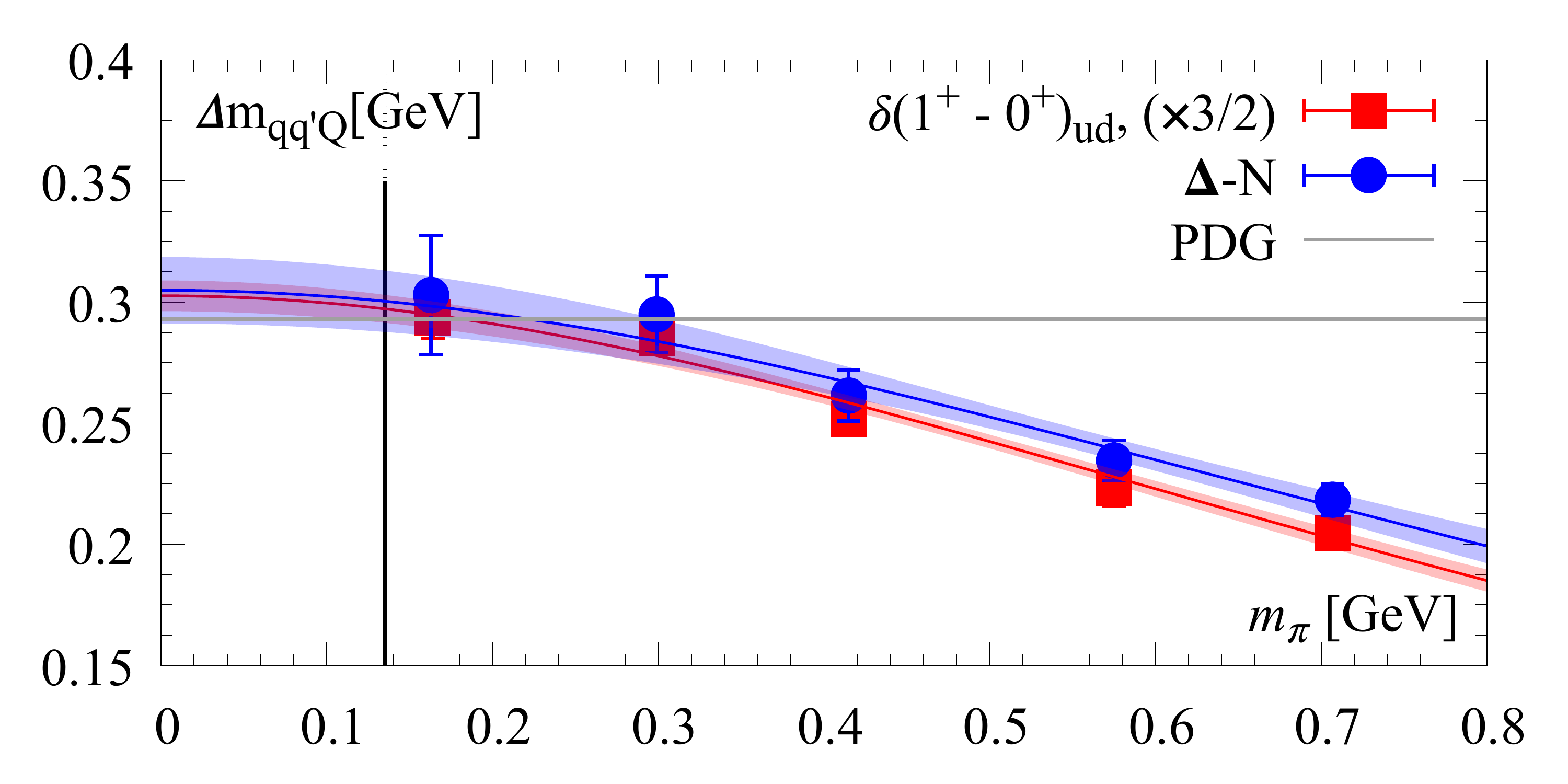}
\caption{{\it Agreement of the bad-good
diquark mass splitting with the 
prediction~\cite{Jaffe:2004ph}, 
$\delta(\Delta-N)=3/2 \times \delta(1^+-0^+)_{ud}$.
}}
\label{fig:spec_panel_supp}
\end{figure}

To study the mass differences between good and 
bad diquarks, shown in top and middle
panels of Fig.~\ref{fig:spec_panel}, 
we fix the energies in the following way:
First, we analyze the smeared and unsmeared correlators separately. For 
each dataset, we consider one- and
two-state fits. The fit window in
Euclidean time is the longest for which
both fits give ground-state energies
consistent within errors. The ground-state
energies of the smeared and unsmeared 
data sets are then averaged, and 
the larger of the two uncertainties
assigned as the final error.

When performing the extrapolations to physical 
$m_\pi$, we also considered combined fits with 
the Ansatz Eq.~(\ref{eq:adiq}) using 
free $n$ and shared $B$, but did not find an 
improvement and therefore quote results from 
individual fits. As a further consistency 
check note that all three extrapolations 
intersect for large $m_\pi$ at the 
$n_f=3$ flavor-symmetric point without 
having enforced this expectation through 
a shared parameter.

In the bottom panel of Fig.~\ref{fig:spec_panel} 
we show the difference in mass between 
single-static octet baryons and static-light 
pseudoscalar mesons, as explained in 
the text. In this data the excited state 
contamination is much larger and we extract 
the masses by fitting both smeared and
unsmeared data with a two-state Ansatz. 
As before we take the values from the 
longest time interval where the fitted 
ground states agree within errors, 
average the results, and quote the 
larger of the two uncertainties as 
our error. For the $ud-u$, 
$\ell s-s$ and $\ell s - \ell$ cases we 
observe the expected $n_f=3$ degeneracy 
as $m_{u,d}\rightarrow m_s$.

As a final diquark spectroscopy
investigation, we compare the bad-good diquark 
mass difference with the $\Delta$-$N$ mass
splitting for each of our 5 ensembles.
This comparison is motivated by the 
observation~\cite{Jaffe:2004ph} that,
in the one-gluon-exchange approximation, and 
chiral limit,
$\delta (\Delta - N) = \frac{3}{2} \delta(1^+ - 0^+)$.
Fig.~\ref{fig:spec_panel_supp} compares the
left- and right-hand sides of this 
relation, the red curve showing the 
appropriately rescaled version of the 
$\delta (1^+ - 0^+)_{ud}$ fit of
the middle panel of 
Fig.~\ref{fig:spec_panel}
and the blue curve a similar fit to the 
$\Delta$-$N$ data with 
$A=\frac{3}{2}\times 0.203(9),~ B=1.10(11)$ GeV.
Agreement between the two is excellent in the
chiral limit, and remains very good over the
whole mass range.
Measurements of the $\Delta -N$ splitting 
were newly performed for this study, 
using the same propagators as for 
the static baryons shown so far. Since 
in this case we do not benefit from 
the cancellation of the heavy 
quark mass, and nucleon correlators
generally suffer from the well known 
signal-to-noise problem, we expect larger
uncertainties in this study. The 
situation is further complicated as 
the $\Delta$ baryon is 
a resonance in nature: a single operator 
analysis, as performed here, can 
capture only its rough features. To stabilise the extraction of the masses
here, we fitted the channel pairs
$\Gamma=(\gamma_i,\sigma_{i0})$ and
$(\gamma_5,\gamma_5\gamma_0)$ 
simultaneously with single exponentials
and chose to quote the parameters 
from the longest combined plateau. 
For the physical-point extrapolation,
the results were fitted to the same 
Ansatz Eq.~(\ref{eq:adiq}), in the same 
way as before.

\section{Lattice structure analysis details}
\label{app:struct}

The diquark size $r_0$ can be estimated from 
the fitted rate of the exponential decay, 
$\sim \exp(-r_{qq'}/r_0)$, of the
density-density correlator 
$\rho_2^\perp (R,r_{qq^\prime})$, with 
$r_{qq^\prime}$ the distance between the 
two current insertion points: see 
Fig.~\ref{fig:rad_panel} (left).

The colored bands (one for each $m_\pi$) 
are the result of performing a combined 
fit for all available $R$ to a single 
exponential with shared size parameter 
$r_0$ and separate amplitudes. Note 
that our lattice spatial 
size is about $5 r_0$, so we neglect 
corrections caused by periodic boundary 
conditions which were studied in 
\cite{Green:2010vc}. We checked the 
dependence of $r_0(R)$ on $R$ through 
individual fits and found no 
significant dependence for $R\in[3:6]$. 
Similar findings were reported in 
\cite{Alexandrou:2005zn}.
Of course, if $R$ is increased beyond 
$\sim 1\, \rm{fm}$, the effective string 
between the static quark and the diquark 
will break and a light baryon will form, 
with qualitatively different diquark 
correlations. Our study does not consider 
this large-distance regime.
Also, for a given $R$, we normally would
quote the number from the largest stable 
fit window. However, due to our chosen 
geometry we may expect interference 
from the static quark when 
$r_{qq'} \gtrsim R$, and we limit the 
fit window accordingly.

\begin{figure}[t!]
\centering
\includegraphics[width=0.55\columnwidth]{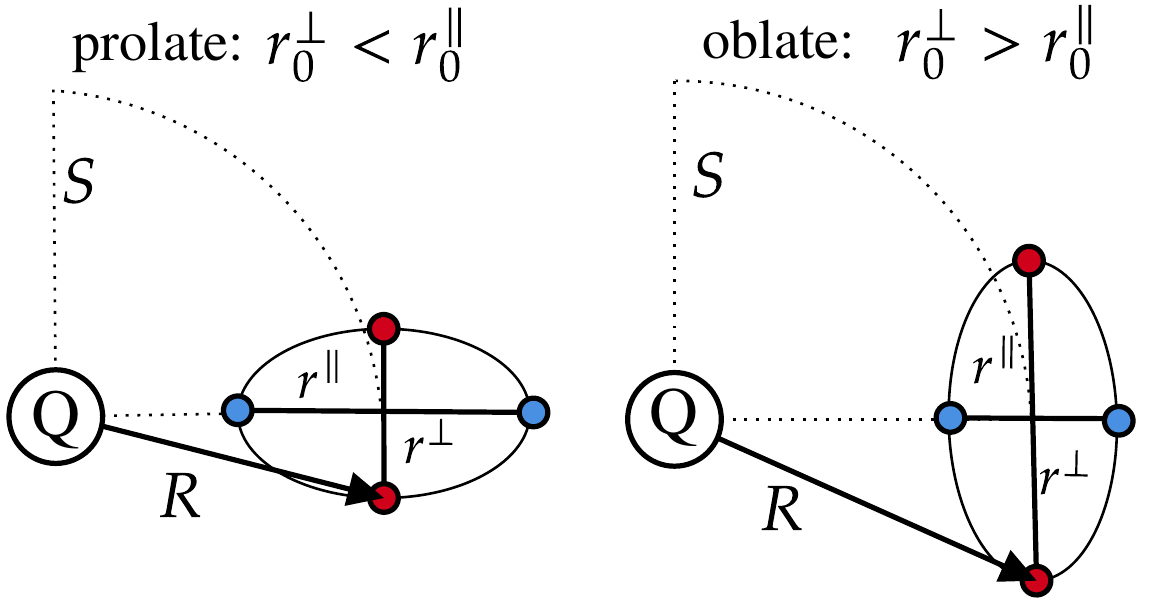}
\caption{{\it
Diquark shape: 2D sketch of the 2 current 
insertions. Comparing exponential 
fall-offs in the $\perp$ and $\parallel$ 
directions gives a measure of the diquark
shape: prolate (left) or oblate (right). 
}}
\label{fig:geom23}
\end{figure}

\begin{figure}[t!]
\centering
\subfloat{\includegraphics[width=0.3\columnwidth]{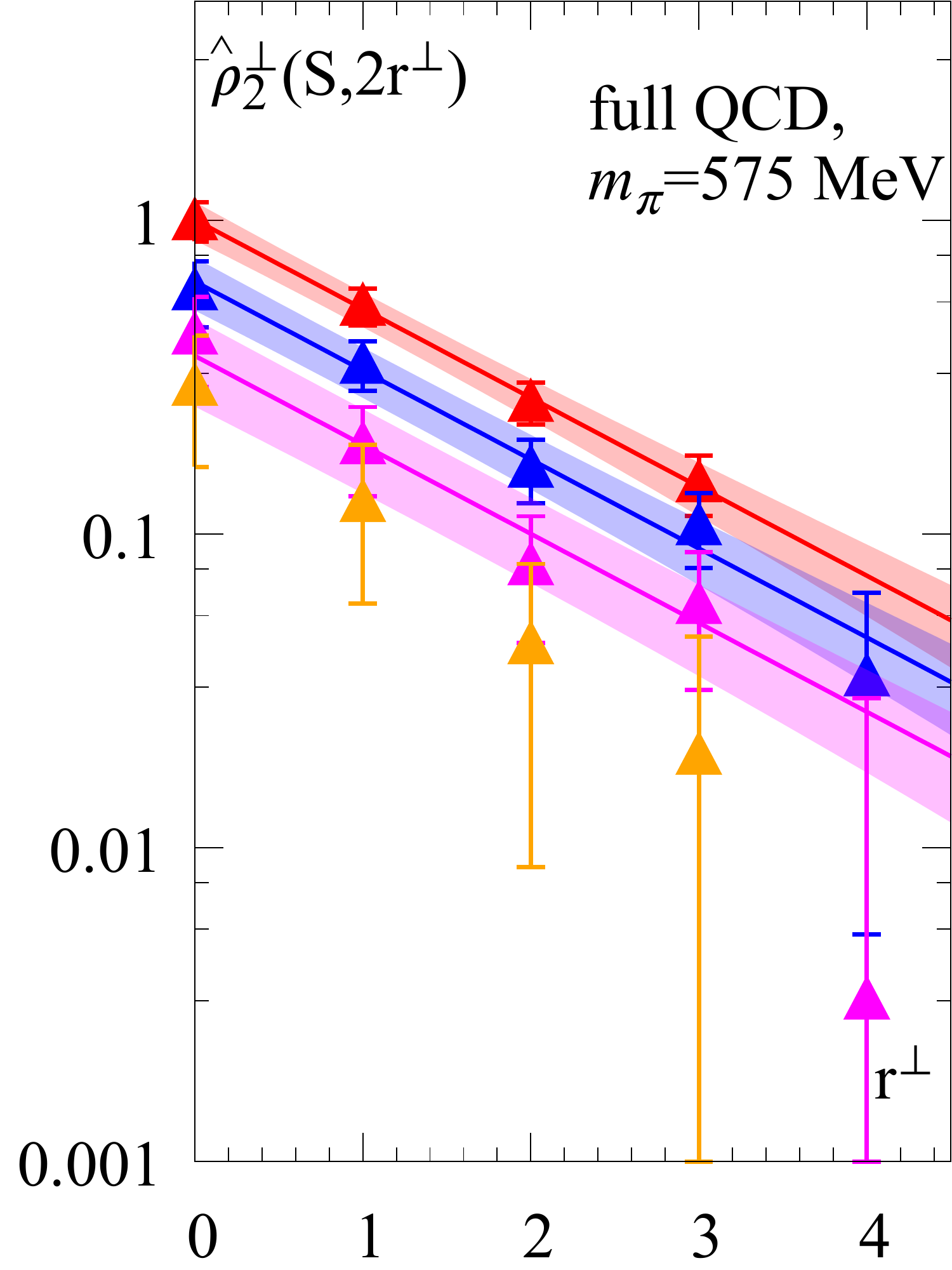}}
\subfloat{\includegraphics[width=0.3\columnwidth]{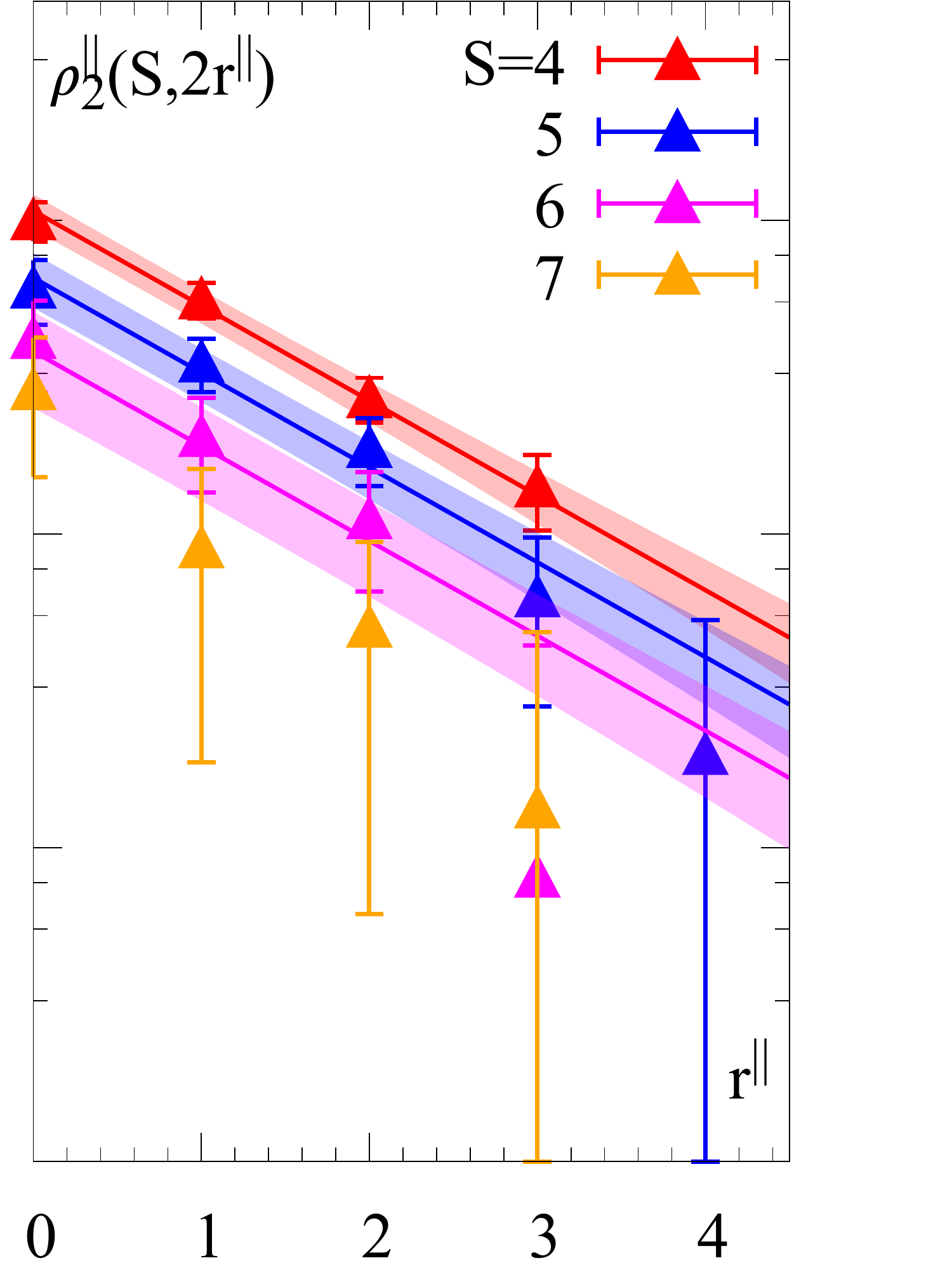}}
\caption{
{\it Good diquark shape. 
$m_\pi = 575$ MeV results for the 
$r_{qq^\prime}$-dependence of 
$\rho_2(r_{qq^\prime},S,\phi )$ for 
tangential ($\phi=\pi /2$, left panel) 
and radial ($\phi =\pi$, right panel) 
quark-quark orientations. The colored
error bands are the results of combined 
fits to data for each of 
$S = 4a$, $5a$ and $6a$. }}
\label{fig:obl_panel_supp}
\end{figure}

Separate sizes $r_0^\perp$ and
$r_0^\parallel$ can be defined for the 
tangential and radial geometries shown 
in Fig.~\ref{fig:geom23}. The ratio of 
sizes $r_0^\perp/r_0^\parallel$ then
gives a measure of whether the diquarks 
are spherical 
($r_0^\perp/r_0^\parallel=1$), 
prolate ($r_0^\perp/r_0^\parallel < 1$), 
or oblate ($r_0^\perp/r_0^\parallel > 1$).

To estimate $r_0^\perp$ and 
$r_0^\parallel$, we measure 
$\rho_2(r_{qq^\prime},S,\phi )$ for the two
geometries of Fig.~\ref{fig:geom23}, with
$r_0^\perp$ and $r_0^\parallel$ 
corresponding to $\phi =\pi /2$ and $\pi$
in Fig.~\ref{fig:geom1}, 
respectively. When the line labelled $S$ in
Fig.~\ref{fig:geom1} points along the $x$ 
axis, the current insertion points for 
the radial configuration (the blue points 
in Fig.~\ref{fig:geom23}) are 
${x}_{1,2}=(S\pm r^\parallel,0,0)$, while
those for the tangential configuration 
(the red points in Fig.~\ref{fig:geom23})
are ${x}_{1,2}=(S,\pm r^\perp,0)$. For 
simplicity, we take the line labelled 
$S$ to always lie in one of
the $x$, $y$ or $z$ axis directions,
considering all such permutations.

Focusing first on the radial case, 
$x_1+x_2=2S$ is constant at fixed $S$ and 
independent of $r^\parallel$. Here we
define our radial size parameter, 
$r_0^\parallel$, at this fixed $S$, 
by fitting the $r^\parallel$ 
dependence to the form
\begin{align}
\rho_2^\parallel (S,r^\parallel ) \sim \exp(-r^\parallel /r_0^\parallel )~~.
\end{align}
Since no obvious $S$ dependence is
observed, we arrive at $r_0^\parallel$ 
by analyzing the data in a combined fit 
for several $S$ using 
the same fit method applied before. 

In the tangential case, a complication
arises. Our previously introduced
size parameter, $r_0$, was
defined through a fit to
$\rho_2^\perp (R,r^\perp )$ with
variable $r^\perp$, but fixed $R$. 
Since, however, 
$R=\sqrt{(r^\perp )^2+S^2}$, 
when $r^\perp$ varies at fixed $R$,
$S$ also varies. This is not the
fixed-$S$ situation used to define 
$r_0^\parallel$. We thus need to
define an alternate tangential
size parameter, $r_0^\perp$,
through a fit to data with variable
$r^\perp$ but fixed $S$, in order to
compare tangential and radial size
parameters both defined at fixed $S$. As
seen above, the density-density correlation
$\rho_2^\perp (R,r^\perp )$ at fixed 
$r^\perp$ varies with $R$. We find 
this dependence well described by an 
exponential form $\sim \exp (-2R/R_0)$. 
The dependence of the density-density
correlation on $r^\perp$ at fixed $S$ in
the tangential configuration can then be 
obtained by fitting the product
$\rho_2^\perp (R,r^\perp ) \exp(+2R/R_0)$,
evaluated at fixed $S$ and variable
$r^\perp$, to the form 
$\exp(-r^\perp /r_0^\perp)$.
Since the fixed-$S$ tangential 
$r^\perp =0$ and radial $r^\parallel =0$
configurations are geometrically 
degenerate, it is convenient to instead 
use the form $\exp(-r^\perp /r_0^\perp)$ 
to fit the modified product 
\begin{align}
\hat{\rho}_2^\perp (S,2r^\perp)
=\rho_2(R,2r^\perp)\,\exp(+\frac{2R}{R_0})\, \exp(-\frac{2S}{R_0})\, ,
\label{app:ans_struct_alt}
\end{align}
with $R_0$ a second fit parameter, and
the right-hand side evaluated at fixed $S$.
The extra $r^\perp$-independent factor, 
$\exp (-2S/R_0)$, ensures that, in the
limit that $r^\perp\rightarrow 0$ and 
hence $R\rightarrow S$, the quantity 
being fit reduces to 
$\rho_2^\perp (R=S, r^\perp =0)$. 
Since this is identical to the analogous
zero-separation quantity,
$\rho_2^\parallel (S,r^\parallel =0)$,
which enters the fit used to determine
$r_0^\parallel$, this choice ensures
a common normalization for the tangential
and radial fits, at the 
$r^\perp =r^\parallel =0$ point
common to both.

In  the left and right panels of 
Fig.~\ref{fig:obl_panel_supp} we show, for
the $m_\pi=575~\rm{MeV}$ ensemble and $S$ 
ranging from $4$ to $7$ times the lattice 
spacing, the dependences of
$\hat{\rho}_2^\perp (S,2r^\perp )$ on 
$r^\perp$ and 
$\rho_2^\parallel (S,2r^\parallel )$ on
$r^\parallel$, respectively.
The results are normalized so 
$\hat{\rho}_2^\perp (S,2r^\perp )$ and
$\rho_2^\parallel (S,2r^\parallel )$ 
take the common value $1$ at $S=4a$ and 
zero separation.

While the parameters $r_0^\perp$ and 
$R_0$ have been determined in the
two-parameter
fit described above, $R_0$ could, in 
principle, also be determined by fitting
$\rho_2^\perp (R,r^\perp )$, with 
$r^\perp$ fixed to zero, to the form 
$\sim \exp (-2R/R_0)$. 
We found that inserting the resulting $R_0$ into the Ansatz Eq.~(\ref{app:ans_struct_alt}) and subsequently fitting $r_0^\perp$ produced no improvement over the direct two-parameter fit result, once errors were propagated and correlations taken into account. 

In both the radial and tangential cases, 
we have propagated all errors within a 
bootstrap procedure, which allows us to 
also evaluate the uncertainty on the 
ratio $r_0^\perp/r_0^\parallel$ 
consistently. Results for this ratio 
are shown as a function of $m_\pi^2$ 
in Fig.~\ref{fig:obl_panel} of the main text. 
We observe that the errors on the data 
limit the precision of the analysis, with 
a stable result, for example, not even 
attainable for the $m_\pi=164~\rm{MeV}$ 
ensemble. This is due in part to the
noisiness of the results at large $S$,
which were not precise enough to constrain
$R_0$, $r_0^\perp$ and $r_0^\parallel$ 
further.

\bibliographystyle{JHEP}
\bibliography{references}

\begin{thebibliography}{10}
\bibitem{renorm}
{The slight difference in the two $0^+$ results originates from their two
  different renormalisation constants. Unfortunately these are not available
  for the ensembles studied.}
\end{thebibliography}

\end{document}